# DIFFICULTIES APPLYING RECENT BLIND SOURCE SEPARATION TECHNIQUES TO EEG AND MEG

K. H. KNUTH
*Dynamic Brain Imaging Laboratory*
*Department of Neuroscience*
*Albert Einstein College of Medicine*
Bronx NY 10461, USA
kknuth@balrog.aecom.yu.edu

**Abstract.** High temporal resolution measurements of human brain activity can be performed by recording the electric potentials on the scalp surface (electroencephalography, EEG), or by recording the magnetic fields near the surface of the head (magnetoencephalography, MEG). The analysis of the data is problematic due to the fact that multiple neural generators may be simultaneously active and the potentials and magnetic fields from these sources are superimposed on the detectors. It is highly desirable to un-mix the data into signals representing the behaviors of the original individual generators. This general problem is called blind source separation and several recent techniques utilizing maximum entropy, minimum mutual information, and maximum likelihood estimation have been applied. These techniques have had much success in separating signals such as natural sounds or speech, but appear to be ineffective when applied to EEG or MEG signals. Many of these techniques implicitly assume that the source distributions have a large kurtosis, whereas an analysis of EEG/MEG signals reveals that the distributions are multimodal. This suggests that more effective separation techniques could be designed for EEG and MEG signals.

## 1  Mixing and Separation of EEG Signals

Propagation and mixing of signals from simultaneous active sources occurs in many physical situations. Probably the most familiar situation is the "cocktail party problem" where there are many speakers, or sources of acoustic signals, and the listener detects mixtures of these signals. While the listener has only two ears, we will consider cases that are much less difficult where there are as many detectors as sources. Often we are searching for a signal embedded in noise of some sort. In cases where much information is known about the signal, such as temporal behavior or frequency content, there are useful standard techniques that can assist in isolating the signal. However, when little specific information is known about the temporal or frequency structure or when signal and noise overlap significantly in the time or frequency domain separation becomes difficult.

This is a serious problem in the interpretation of electrophysiological data. During a given experiment there are many simultaneous sources of electrical activity in the brain and the currents generated by all of these sources propagate throughout the head and are linearly superimposed on one another. Thus any detector of electromagnetic activity in the brain will pick up a linear combination of all of the active sources.

In the experimental situation there are $m$ detectors, which are used to record the electric potentials or magnetic fields generated by synchronous neural activity in the brain. These detectors are either electrodes, which are placed on the scalp for electroencephalography (EEG) or superconducting quantum interference devices (SQUIDs) coupled to coils placed near the head for magnetoencephalography (MEG). If during the recording there are $n$ active neural generators then mixtures of these $n$ signals are received by the $m$ detectors. This linear instantaneous mixing process can be represented in terms of a matrix multiplication between an $m \times n$ mixing matrix and the $n$-dimensional source vector $\mathbf{s}(t)$. The $m$-dimensional signal, $\mathbf{x}(t)$, can be represented by

$$\mathbf{x}(t) = \mathbf{A}\, \mathbf{s}(t). \qquad (1)$$

The most obvious way to separate these signals is to find the matrix $\mathbf{A}^{-1}$; however, there is actually more freedom present in the problem. The order of the sources is not important and so any permutation of the sources is also considered a solution. In addition, since both $\mathbf{A}$ and $\mathbf{s}(t)$ are unknown, the scaling of the sources is indeterminate and we therefore seek an unmixing matrix, $\mathbf{W}$ of the form:

$$\mathbf{W} = \Sigma \Pi \mathbf{A}^{-1}, \qquad (2)$$

where $\Sigma$ is a diagonal scaling matrix, and $\Pi$ is a permutation matrix. The resulting unmixed signals can then be expressed as

$$\mathbf{u}(t) = \mathbf{W}\, \mathbf{x}(t). \qquad (3)$$

## 2  Summary of Blind Source Separation Techniques

In the last several years there has been much work on the problem of blind source separation, which has resulted in many diverse approaches. Most of these approaches use higher-order statistics, minimum mutual information, and maximum entropy in their solutions. In addition to the maximum entropy solution, which is relevant to this discussion, we will mention a higher-order extension to the more popular principal component analysis.

Principal component analysis (PCA) is a technique that has been used to attempt to separate EEG signals into their principal, or decorrelated, components. It is based on the fact the cross-cumulants of statistically independent signals are identically zero. In PCA one diagonalizes the covariance matrix, which results in the decorrelation of the signals. Because this technique deals only with the covariance, it takes into account only the second-order statistics and disregards information represented by the higher-order statistics. In addition, by using only second-order statistics to describe the signals one implicitly assumes that the sources are Gaussian, which is often inappropriate. Algebraic techniques utilizing the information present in the higher-order statistics exist. The most direct technique relies on the diagonalization of the fourth-order tensor derived from the fourth-order cross-cumulants (Cardoso 1995). Other similar techniques exists where one minimizes functions expressed in terms of higher-order moments or cumulants of the input signals, which represent series expansions of the mutual information (Comon 1994; Yang & Amari 1997).

The term independent component analysis (ICA) refers to finding the statistically independent sources responsible for the data. The approach developed by Bell and Sejnowski (1995) uses a model where the mixed input is presented to a neural network that uses a sigmoidal nonlinearity to transform the input into the output. The weights are then varied so that the joint entropy of the output is maximized. In maximizing the joint entropy of the output one expects to minimize the mutual information between the outputs thus ensuring their statistical independence.

The maximum likelihood approach uses a latent variable model of the sources and finds a mixing matrix that maximizes the likelihood of the model given the data. It has been shown (MacKay 1996; Pearlmutter & Parra 1996; Cardoso 1997) that the maximum likelihood technique is equivalent to ICA where the models of the probability distributions of the source amplitudes are represented by the sigmoidal nonlinearites in Bell and Sejnowski's neural network. The maximum likelihood approach is essentially equivalent to the Bayesian approach, except that in the former the assumption is implicitly made that the mixing matrix prior is a constant.

## 3   Derivation of the Bell-Sejnowski Algorithm

Here we present a derivation of the Bell-Sejnowski algorithm starting with Bayes Theorem (a similar derivation can be found in MacKay 1996). The main purpose of this derivation is to elucidate the assumptions that result in the algorithm. We use a model where the data, $\mathbf{x}(t)$, is generated by a linear transformation of the source behavior, $\mathbf{s}(t)$, as described in eqn. (1). The first simplifying assumption is that the mixing matrix, $\mathbf{A}$, is square, i.e. the number of detectors, $m$, equals the number of sources, $n$. This greatly simplifies the formulation of the problem since the inverse of a non-singular square matrix is well defined. To relax this assumption would most likely require the use of the pseudoinverse of the matrix $\mathbf{A}$ (Penrose 1955). In an experimental setting, we usually have a situation where the number of detectors is much greater than the number of the sources, and intuition would suggest that this extra information would increase our ability to infer the source behavior.

Let $\mathbf{x}$ represent the $n$-dimensional vector of mixtures, and denote an individual mixture as $x_i$, where $x_i$ is understood to be the time-series data received by detector $i$. Likewise denote the $n$-dimensional vector of sources as $\mathbf{s}$, and an individual source as $s_i$, where $s_i$ is understood to be the time-series representing the signal emitted by source $i$.

Then the probability of the mixing matrix and the sources, given the data and any prior information, *I*, is given by Bayes Theorem:

$$P(\mathbf{A}, \mathbf{s} | \mathbf{x}, I) \propto P(\mathbf{x} | \mathbf{A}, \mathbf{s}, I) \cdot P(\mathbf{A} | I) \cdot P(\mathbf{s} | I). \tag{4}$$

Since we do not know much about the mixing matrix, A, we can indicate our ignorance by assigning a flat prior that is constant for all appropriate matrices,

$$P(\mathbf{A} | I) = \text{constant}, \tag{5}$$

resulting in

$$P(\mathbf{A}, \mathbf{s} | \mathbf{x}, I) \propto P(\mathbf{x} | \mathbf{A}, \mathbf{s}, I) \cdot P(\mathbf{s} | I). \tag{6}$$

Since there is a relationship between the mixing matrix, **A**, the sources, **s**, and the mixtures, **x**, we do not need to find both **A** and **s**. We can simplify the problem by treating **s** as a nuisance parameter and marginalizing over **s** by integrating the posterior probability with respect to **s**:

$$P(\mathbf{A} | \mathbf{x}, I) \propto \int d\mathbf{s}\, P(\mathbf{x} | \mathbf{A}, \mathbf{s}, I) \cdot P(\mathbf{s} | I). \tag{7}$$

Now we make the assumption that the sources are statistically independent so that *P*(**s**|*I*) can be represented as the product of the probabilities of the independent sources, $p_j(s_j)$. In addition, assuming the noiseless case, we can write the likelihood of **x** as a product of $\delta(x_i - A_{ik} s_k)$ for each of the mixtures, where the Einstein summation convention is used to denote the matrix multiplication:

$$P(\mathbf{A} | \mathbf{x}, I) \propto \int d\mathbf{s} \prod_i \delta(x_i - A_{ik}\, s_k) \cdot \prod_l p_l(s_l) \tag{8}$$

$$\propto \frac{1}{\det \mathbf{A}} \prod_l p_l\!\left(A^{-1}_{lk}\, x_k\right) \tag{9}$$

Looking at the logarithm of the probability we get

$$\log P(\mathbf{A} | \mathbf{x}, I) = -\log \det \mathbf{A} + \sum_l \log p_l\!\left(A^{-1}_{lk}\, x_k\right) + C \tag{10}$$

where the constant, C, is the logarithm of the proportionality constant implicit in eqn. (9). The results are simplified if we write **W** ≡ **A**$^{-1}$:

$$\log P(\mathbf{A} | \mathbf{x}, I) = \log \det \mathbf{W} + \sum_l \log p_l\!\left(W_{lk}\, x_k\right) + C \tag{11}$$

To obtain the best estimate of the mixing matrix **A** we want to find the maximum of the log of the posterior probability, $P(\mathbf{A}|\mathbf{x},I)$. We proceed by taking the derivative of the log of the posterior probability with respect to the matrix elements of **W**. We define as in eqn. (3), $u_i = W_{ij} x_j$.

$$\frac{\partial}{\partial W_{ij}} \log P(\mathbf{A}|\mathbf{x},I) = \frac{\partial}{\partial W_{ij}} \left[ \log \det \mathbf{W} + \sum_l \log p_l(W_{lk} x_k) + C \right] \quad (12)$$

$$= A_{ji} + \frac{\partial}{\partial W_{ij}} \left[ \sum_l \log p_l(W_{lk} x_k) \right] \quad (13)$$

$$= A_{ji} + \frac{\partial u_i}{\partial W_{ij}} \frac{\partial}{\partial u_i} \left[ \sum_l \log p_l(u_l) \right] \quad (14)$$

$$= A_{ji} + x_j \frac{\partial \log p_i(u_i)}{\partial u_i} \quad (15)$$

$$= A_{ji} + x_j \left( \frac{p'_i(u_i)}{p_i(u_i)} \right)_i \quad (16)$$

This can be expressed in matrix form

$$\frac{\partial}{\partial \mathbf{W}} \log P(\mathbf{A}|\mathbf{x},I) = \mathbf{W}^{-T} + \left( \frac{p'_i(u_i)}{p_i(u_i)} \right)_i \mathbf{x}^T, \quad (17)$$

where the ratio of the probability density of $u_i$ to the probability distribution of $u_i$ is a column vector.

Due to the high dimensionality of the space, ($n \times n$), we search for the maximum of the log of the posterior probability using a stochastic gradient method. We begin by making an initial guess as to the value of **W**, and ascend the gradient while updating our estimate by adding a term proportional to the derivative in eqn. (17):

$$\mathbf{W}_{i+1} = \mathbf{W}_i + l \, \Delta \mathbf{W} \quad (18)$$

where $\Delta \mathbf{W}$ is $\frac{\partial}{\partial \mathbf{W}} \log P(\mathbf{A}|\mathbf{x},I)$ as in eqn. (17), and $l$ is the proportionality constant or learning rate. However, the equation above is not covariant since **DW** is not actually a matrix, but is a gradient of a scalar field, or a one-form. The above equation can be made covariant by post-multiplying the gradient by $\mathbf{W}^T\mathbf{W}$ (Amari et al. 1996; MacKay 1996) resulting in:

$$\Delta \mathbf{W} = \mathbf{W} + \left(\frac{p'_i(u_i)}{p_i(u_i)}\right)_i \mathbf{u}^T \mathbf{W}. \qquad (19)$$

The data is used to provide an unbiased estimate of the probability distribution and density above, and the algorithm is iterated until it converges on a solution. This result is identical to the Bell & Sejnowski neural network implementation of the ICA algorithm.

The advantage of the Bayesian treatment over other approaches is that the formalism makes explicit the assumptions that go into the mathematical description of the problem. In our example we assume nothing about the form of the mixing matrix, which is made explicit by the fact that the prior $P(\mathbf{A}|I)$ is uniform. We assumed that the data is noise-free which was made explicit by the use of a delta function to express the likelihood of the data given the model. Finally, the sources were assumed to be independent which is represented by the fact that we wrote the joint source prior as a product of the individual source priors, $p_i(u_i)$, whose forms can then be explicitly described.

As we will show, the choice of the source prior can be a delicate matter. In ICA the most commonly used priors are the typical threshold functions used in neural models such as the sigmoid, hyperbolic tangent, and arctangent. If one can estimate the mean and variance of the sources, we may be tempted to use a Gaussian MaxEnt prior. However, use of the Gaussian prior can be shown to only diagonalize the covariance matrix of the data, which results in an algorithm akin to principal component analysis. Another notable approach (Pearlmutter and Parra 1996) uses a more elaborate prior that amounts to a sum of sigmoid functions.

In the case that there is noise present in the data, one can replace the delta function representation of the likelihood in eqn. (8) with a Gaussian distribution:

$$P(\mathbf{x}|\mathbf{A},\mathbf{s},I) = \prod_i \frac{1}{\sqrt{2\pi}\sigma} Exp\left[\frac{(x_i - A_{ik}s_k)^2}{2\sigma^2}\right] \qquad (20)$$

where $\sigma$ is the variance of the noise. In general this technique is useful, but depending on the form of the source priors the integrals may not be analytically solvable. In the case of EEG or MEG data, event-related potentials have been averaged over many stimuli and the variance of the noise in the channels can be estimated from the pre-stimulus portion of the recordings. This variance could be used in conjunction with eqn. (20) above to account for the noise in EEG or MEG data.

## 4 Demonstration of the Bell-Sejnowski Algorithm using Audio Signals

In this section we apply the algorithm to a set of artificially mixed audio signals. The set is comprised of sounds from the television show "Star Trek" and is composed of four speech sounds from four different speakers and two artificial sounds, a hail (whistle), and a photon torpedo blast (a noisy frequency glide). No artificial noise has been added to the signals. In keeping with Bell and Sejnowski's original implementation, we choose a sigmoid function as the source prior distribution (Figure 1)

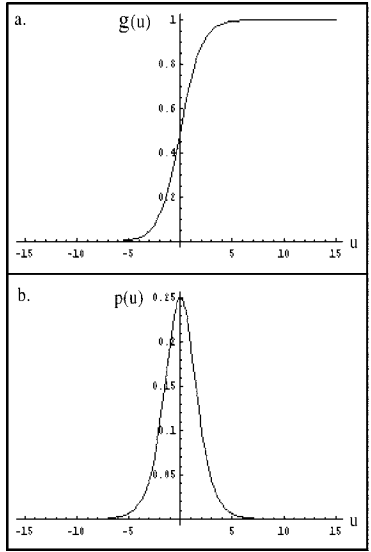

$$g(u) = \frac{1}{1+e^{-u}}, \qquad (21)$$

so that the source prior density is

$$p_i(u_i) = \frac{dg(u_i)}{du_i}. \qquad (22)$$

*Figure 1. (a) The sigmoidal probability distribution, (b) and its probability density.*

By choosing this source prior we are assuming that the probability density functions (p.d.f.s) of the sources are more sharply peaked than a Gaussian density, i.e. the source p.d.f.s have a high kurtosis.

Figure 2 shows the original signals on the left, the artificially mixed signals in the middle and the unmixed signals on the right. Note that the speech sounds are completely unmixed, although permuted, whereas the other two sounds (the "Hail" and the "Photon Torpedo") are still mixed with one another (Unmix 1 and Unmix 3). This pathological solution is called a diagonal solution, and it can be shown to consist of the sum and difference of re-scaled versions of the two source waveforms, as represented by the matrix multiplication:

$$\begin{pmatrix} u_1 \\ u_2 \end{pmatrix} = \begin{pmatrix} 1 & 1 \\ -1 & 1 \end{pmatrix} \begin{pmatrix} a & 0 \\ 0 & b \end{pmatrix} \begin{pmatrix} s_1 \\ s_2 \end{pmatrix}. \qquad (23)$$

The original waveforms, up to a scale factor, can be restored by taking the sum and the difference of the two diagonal solutions. Interestingly, the incorrect diagonal solutions have p.d.f.s (Figure 3) which better match the source prior in equation (22). The diagonal solution is well described by Bell & Sejnowski (1995), and it is easily visualized by considering two uniformly distributed sources. By taking the sum and difference of two uniformly distributed sources one obtains mixtures that have triangular p.d.f.s that more closely match a peaked density like the sigmoidal source prior. This is an excellent example of how the choice in the source prior can affect the results of the separation.

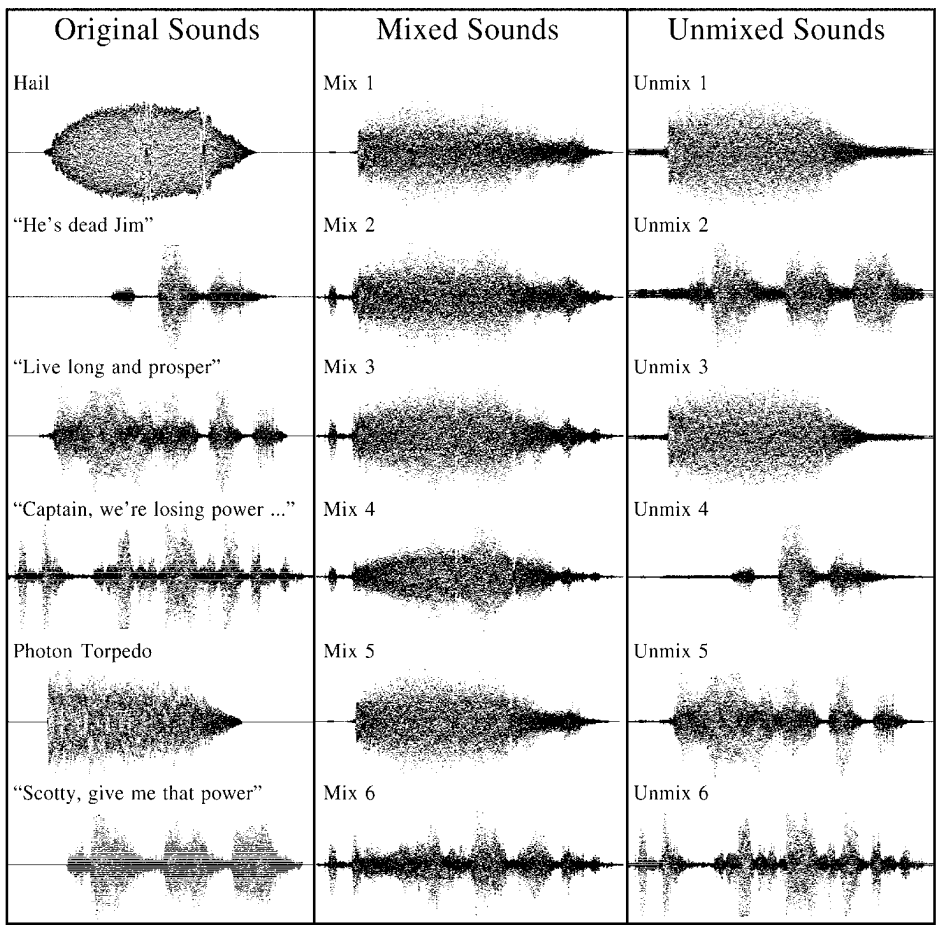

*Figure 2. (Left) Original source waveforms. All are speech sounds except the Hail and the Photon Torpedo, which are a whistle and a noisy frequency glide. (Center) Six distinct mixtures of the original sources used to test the ICA algorithm. (Right) The six waveforms unmixed by the ICA algorithm. Note that the speech sounds are completely unmixed, however, the Hail and the Photon Torpedo are not (Unmix 1 and Unmix 3). These waveforms, Unmix 1 and Unmix 3, are interesting because they represent the sum and difference of the original sources.*

## 5  Application of the Technique to EEG and MEG Signals

The electric and magnetic signals detected by EEG and MEG are generated by highly correlated activity in the brain involving hundreds of thousands of synchronously active neurons. The signals are lower in frequency than are audio signals, with a bandwidth ranging from about 1-100 Hz. Thus the EEG/MEG signals exhibit less variation than audio signals and are much less rich in appearance.

One common technique is the acquisition of event related potentials (ERPs) in which one records electrical events time-locked to a sensory stimulus. By repeatedly presenting the stimulus and averaging the electrical responses, one obtains waveforms consisting of mixtures of signals correlated to the stimulus. These signals represent the average activity from the various neural generators that are evoked by the stimulus. Application of these source separation techniques to EEG and MEG signals would be advantageous. By isolating the responses from different neural generators one could not only accurately locate the anatomical origin of the signals, but also study the dynamical behavior of the generators. In addition, one could also separate the noise from the signal without applying Fourier techniques that can distort the data.

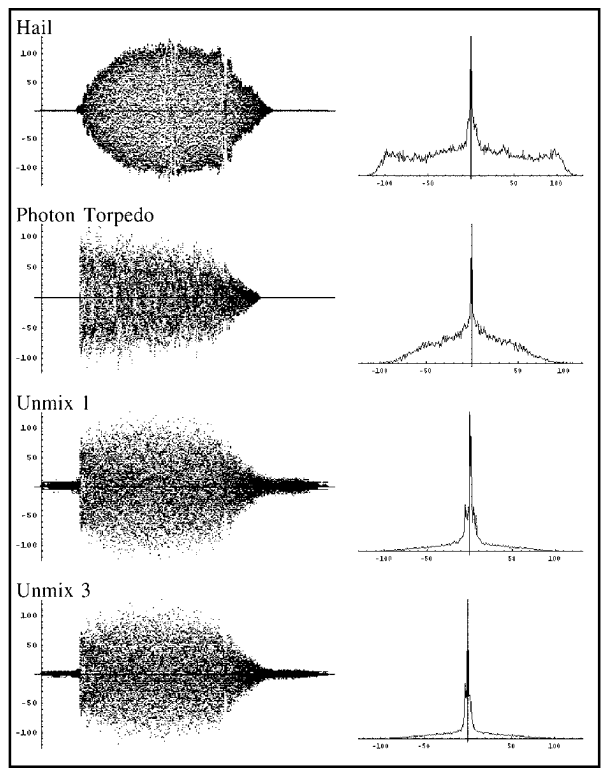

*Figure 3. (Left) The Hail, Photon Torpedo, and their diagonal solutions, which represent the sum and difference of the original waveforms. (Right) Histograms of the amplitudes of the waveforms, which represent their probability density functions. Note that the diagonal solutions have p.d.f.s that better match the source prior density. In addition, the large amount of baseline noise also assists in matching the p.d.f.s of the diagonal solutions with the density assumed by the source prior.*

We now demonstrate the application of the algorithm on an auditory event related response recorded using MEG. The response was evoked using a 0.5 ms square-wave condensation click presented to the right ear. The radial component of the magnetic field was recorded over the left hemisphere of the subject's head using a 37-channel biomagnetometer. Responses were averaged 512 times to extract the signals that are time-locked to the stimulus. The data was sampled at a sampling rate of 1041.7 Hz, and was conditioned between 0.1 and 400 Hz by using digital high and low pass filters. It can be shown that filtering does not affect the mixing matrix; this is because filtering is a linear operation that acts on each channel identically. This particular data set was chosen because the stimulus delivery was time-locked to a 60 Hz cycle, which prevented the 60 Hz line noise from being cancelled out during averaging. This 60 Hz signal provides us with an extra source, which can be used to evaluate the success of the technique. By applying singular value decomposition to the data we estimated that there were about 7 sources responsible for the data. Greater and fewer numbers of sources were tested without an increase of success.

The left side of Figure 4 shows 7 of the 37 original recorded waveforms. These waveforms were selected because they spanned the set of typical waveforms observed across the detector. Physiologically significant peaks in the responses can be seen clearly in the waveform recorded from channel 35. The two small peaks between 0 and 0.1 s are the magnetic equivalent of the electrical middle latency response. The large negative peak just after 0.1 s is known as the M100, and two similar peaks can be seen at 0.2 s and 0.3 s. Channels 5, 11, and 34 show a preponderance of 60 Hz noise, which is known to come from a source independent of the recorded brain activity.

The right side of Figure 4 shows the waveforms that result from applying the ICA algorithm to the data. At first glance the waveforms seem to be relatively independent of one another, however, one can see several peaks that are replicated throughout many of the "unmixed" waveforms. One would also expect that the "unmixed" waveforms should have less noise than the original waveforms and this is obviously not the case. Examination of the power spectra of the "unmixed" waveforms reveals that there is still a large amount of 60 Hz noise in "unmixed" waveforms 2, 5, 6, and 7 demonstrating that the 60 Hz noise has not been extracted from the data.

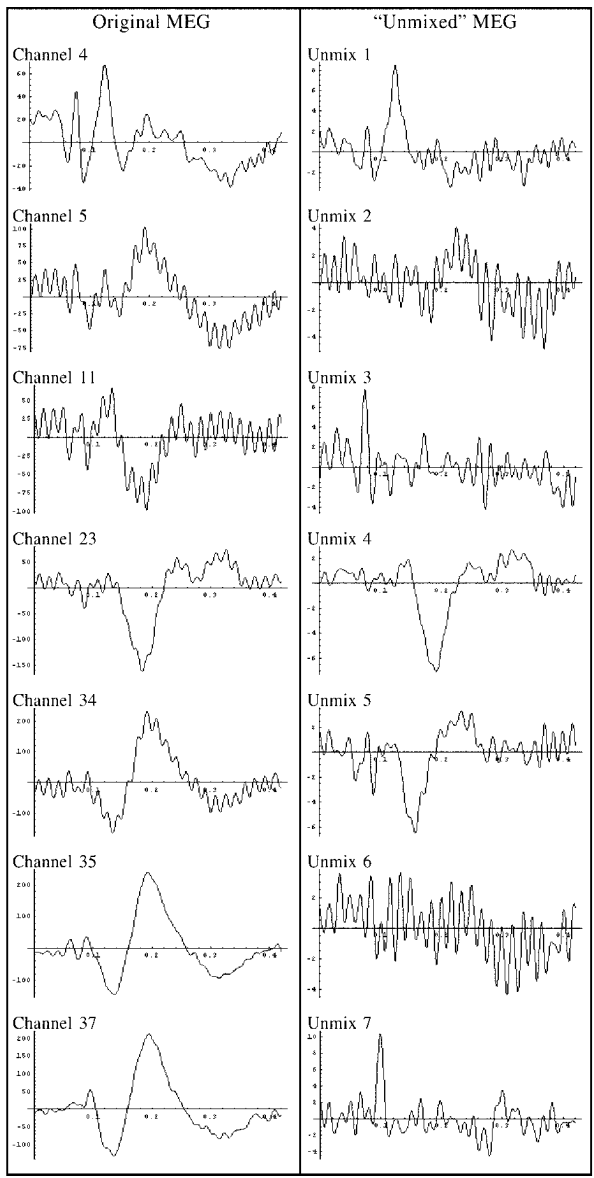

*Figure 4. (Left) Original MEG signals (auditory evoked fields). Note the 60 Hz noise in channels 5, 11, and 34. (Right) "Unmixed" MEG signals. At first glance the waveforms may appear to be independent, but several peaks are still replicated throughout the waveforms. In addition, there is still a prevalence of 60 Hz noise in the "Unmixed" waveforms 2, 5, 6, and 7.*

## 6    Probability Density Functions of Various Signals

A better appreciation for the blind source separation problem can be obtained by examining the p.d.f.s of some of the signals in the examples above. Figure 5 shows waveforms represented by time-series data and the histograms of their amplitudes, which approximate the p.d.f.s of the waveforms. The first waveform shown is Gaussian noise. Contrast this Gaussian p.d.f. with the p.d.f. of the spoken phrase, "Live long and prosper", which has a high kurtosis, visible from the sharp central peak and heavy tails. The broad, flat tails of the p.d.f. of the "Hail" waveform was the cause of the pathological solution described above. From these three examples and the discussion on diagonal solutions one can see how the sigmoid prior is well adapted for separating speech sounds, yet fails to separate sources with broad p.d.f.s.

The following waveforms are more similar to the EEG/MEG signals of interest. The sine wave is a good example as it can be shown analytically that its p.d.f. is a scaled secant function with peaks at ±1. These peaks come from the fact that the derivative of the waveform is zero at amplitudes of ±1. Another illustrative example is the Bessel function, $J_2(x)$. This waveform is qualitatively more similar to what we would expect from a neural source. Again it demonstrates that the p.d.f. will have peaks at amplitudes where the derivative of the waveform is small. These multi-modal distributions have a low kurtosis and it can be shown that the ICA algorithm as such cannot properly separate them.

The last two waveforms are of EEG and MEG data from the experiment described above. It is important to remember that these waveforms are mixtures of sources and thus the p.d.f.s are derived from a convolution of the p.d.f.s of the original sources. With this in mind the overall Gaussian-like shape of the p.d.f.s is not surprising since we expect the Central Limit Theorem to be at work. In addition, the strong spikes in the p.d.f. are due to the spikes in the original source p.d.f.s. This is especially obvious in the MEG waveform where there is sinusoidal 60 Hz noise. The convolution of the secant function, representing the p.d.f. of the noise source, with the MEG source p.d.f.s results in a p.d.f. with a series of strong spikes. Figure 6 shows the p.d.f.s of the first three "unmixed" MEG waveforms. Note that even waveforms 1 and 3, which seem to show rather distinct source-like behavior, have p.d.f.s that are to be expected from mixtures.

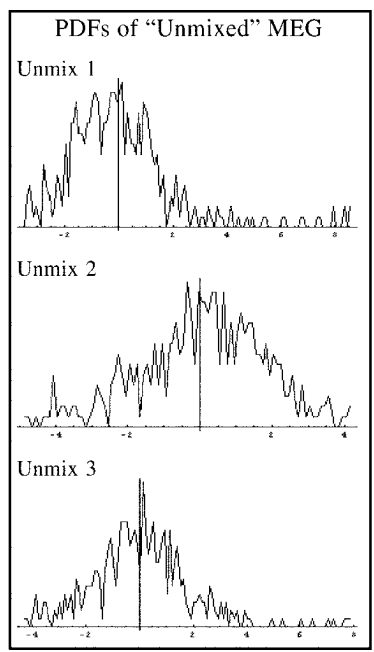

*Figure 6. Histograms of the amplitudes of the first three "Unmixed" MEG waveforms. Note the spiky Gaussian-like shape that fits the assumed source prior density.*

The fact that one expects mixtures to have Gaussian-like p.d.f.s explains why the simple technique of PCA fails to separate EEG signals. In addition, one can see that in the context of ICA, where the source prior densities have a high kurtosis, one also expects mixtures of EEG/MEG sources to better fit the source priors. This effect is similar to the pathological diagonal solutions that result when sources have broad p.d.f.s.

## 7 Discussion

It is often true in Bayesian analysis that the assignment of the prior probabilities is not a critical issue, and this is demonstrated by the impressive ability of ICA to separate speech sounds. Indeed, it is conjectured that the typical source priors utilized in ICA, such as the sigmoid, hyperbolic tangent, and arctangent functions, are sufficient to separate any sources with large kurtosis. A partial proof of this conjecture is given by MacKay (1996) for the case of two sources having the same high kurtosis distributions. Pathological solutions, however, are known to exist for cases where the source p.d.f.s are broad compared to the sharply peaked source prior p.d.f.. We have demonstrated that the separation of EEG and MEG signals is not possible using sharply peaked source priors, and these results suggest that multiply peaked source densities may need to be utilized to obtain proper separation.

It is important to focus on several of the assumptions made in the derivation and application of this technique. Some of these assumptions facilitate the mathematical description of the problem, while others reflect our ignorance about the nature of the signals and the sources.

Three assumptions deal with the mathematical description of the problem. First, we assume the same numbers of sources as detectors. It is expected that with more detectors we have more information about the sources and separation would be facilitated. In this case, however, we are no longer dealing with square mixing matrices, and we would need to develop a solution using pseudoinverse matrices. Second, there is a difficulty in estimating the number of sources. One of the most popular techniques is to use singular-value decomposition, but it is known that the technique is not robust to noise. The advantage of the Bayesian approach here again is that the ratio of probabilities obtained from models using different source numbers can be computed allowing us to choose the optimal model. Third, we have demonstrated the necessity for finding generic source priors that are effective in separating signals with multimodal p.d.f.s. The high-kurtosis source priors that are typically used in ICA result in producing mixtures rather than sources. Generic source priors that can accommodate multimodal p.d.f.s will most likely require many adjustable parameters. Pearlmutter and Parra (1996) suggest using sums of sigmoid functions. However, it is often the case that algorithms become unstable as the number of parameters to be fit becomes large.

We briefly address another assumption that deals with a more difficult issue, that of statistical independence of the sources. Physiologically, strong dynamical coupling is likely to exist between many sources, and this may pose a problem for the separation algorithm. Although these algorithms can often separate sources exhibiting some dependence, this is probably a matter of degree. A more difficult case would be one where a spatially extended active region exhibits spatially-varying dynamical behavior. This is a continuous generalization of the case of discrete coupled sources, and brings into question the validity of the notion of neural generators.

## Acknowledgements

I would like to thank the following people for their assistance, advice, and support: John Broadhurst, Steve Cobb, Joslyne Foley, Irv Hochberg, Lacey Kurelowich, Barry Schwartz, Patti Quint, and Herb Vaughan. This work was supported in part by Albert Einstein College of Medicine, Biomagnetic Technologies Inc., City University of New York Graduate Center, Scripps Research Institute, and NIH NIDCD 5 T32 DC00039-05.

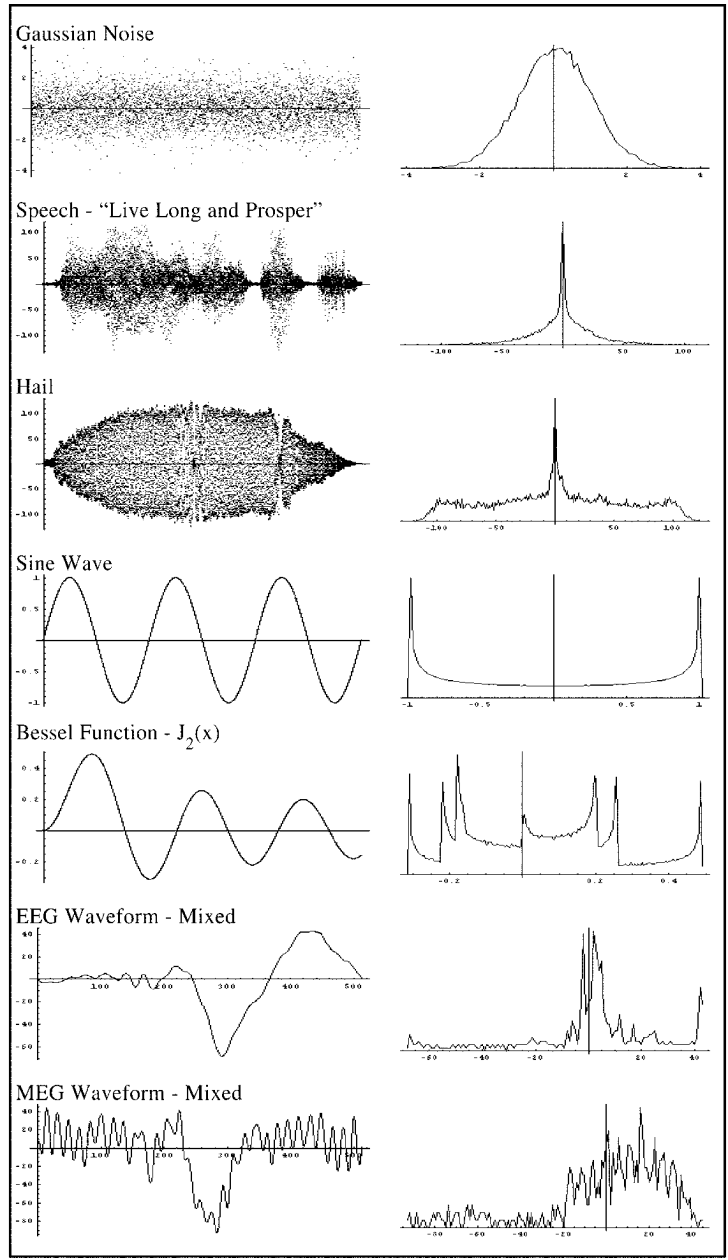

*Figure 5. (Left) Various signals represented by time-series data. (Right) Histograms of the amplitudes of the corresponding time-series. In the limit these histograms approach the probability density functions of the signals' amplitudes. (The scale on the vertical axis is absent for clarity. It is understood that the areas are normalized to unity.)*